\documentclass[10pt,aps,prl,twocolumn,superscriptaddress]{revtex4-2}
\usepackage{psfrag}
\usepackage{mathrsfs}
\usepackage{amssymb,bm}
\usepackage{amsmath,amsthm}
\usepackage{epstopdf}
\usepackage{enumerate}
\usepackage{color}
\usepackage{mathrsfs}
\usepackage{graphicx}
\usepackage{bm}
\usepackage[english]{babel}
\usepackage[T1]{fontenc}
\allowdisplaybreaks
\usepackage{tikz}
\newcommand{\be}{\begin{equation}}
\newcommand{\ee}{\end{equation}}

\newcommand{\red}{\color{red}}

\definecolor{purple}{rgb}{1,0,1}
\definecolor{lime}{HTML}{a6CE39} % needs xcolor

\usepackage{proofread}
\usepackage{comment}
\begin{document}

\title{Thermal origin of the attractor-to-general-relativity in scalar-tensor gravity}

\author{Valerio Faraoni}%\orcidValerio\!} 
\email[]{vfaraoni@ubishops.ca}
\affiliation{Department of Physics \& Astronomy, Bishop's University, 2600 College Street, Sherbrooke, Qu\'ebec, Canada 
J1M~1Z7}

\author{Andrea Giusti}%\orcidAndrea}\! 
\email[]{agiusti@ubishops.ca}
\affiliation{Department of Physics \& Astronomy, Bishop's University, 2600 College Street, Sherbrooke, Qu\'ebec, Canada 
J1M~1Z7}
\affiliation{Department of Physics \& Astronomy, University of Sussex, Brighton, BN1~9QH, United Kingdom}

\begin{abstract}

The convergence of scalar-tensor gravity to general relativity, or the departure from it, are described in a new analogy with heat dissipation in a viscous fluid. This new thermal picture is applied to cosmology, shedding light on  whether gravity  deviates from general relativity early on and approaches it later in  the cosmic history. 

\end{abstract}

%\date{\today}

\maketitle

\noindent {\it Introduction}---General relativity (GR) is quite successful but 
has its problems. First, it predicts spacetime singularities in cosmology and inside 
black holes, where physical quantities diverge, i.e., it predicts its 
own failure. It is believed that quantum mechanics,  at or below the Planck energy scale, will cure these singularities. However, no 
quantum gravity model is satisfactory and there is virtually no 
experiment to guide theory development. The lowest-order quantum corrections to GR 
inevitably modify it by introducing extra degrees of freedom or higher order field 
equations. Therefore, from the point of view of theoretical physics, GR cannot be the 
ultimate theory of gravity. Second, the standard cosmological model, the GR-based $\Lambda$-Cold Dark Matter  model described by the Friedmann-Lema\^itre-Robertson-Walker (FLRW) geometry, is plagued by  various  
tensions. Moreover, in order to explain the present acceleration of the cosmic 
expansion discovered in 1998 with Type Ia supernovae, one must introduce an {\em ad 
hoc} dark energy of mysterious origin akin to a fudge factor. Hence 
researchers have turned to the alternative of modifying gravity at the largest 
(cosmological) scales while giving up dark energy entirely.  The most popular alternative to GR is $f(R)$ gravity (where $R$ is the Ricci scalar curvature 
and $f$ is a non-linear function). A proof 
of principle was given that $f(R)$ gravity can explain the cosmic acceleration without 
dark energy (see \cite{Sotiriou:2008rp,DeFelice:2010aj,Nojiri:2010wj} for reviews).

The simplest class of theories extending GR is scalar-tensor (ST) gravity, which only adds an extra scalar propagating degree of freedom $\phi$ to the GR degrees of freedom 
(the two massless, spin two modes appearing as  
gravitational wave polarizations). This extra scalar field $\phi$ corresponds, roughly speaking, to 
the inverse of the effective gravitational coupling strength $G_\mathrm{eff} \simeq \phi^{-1}$. 
That is, in ST gravity Newton's constant $G$ becomes a dynamical field. This 
was the original Brans-Dicke proposal \cite{Brans:1961sx}, later extended to more 
general theories \cite{Bergmann:1968ve,Nordtvedt:1968qs,Wagoner:1970vr,Nordtvedt:1970uv}. Notably,  $f(R)$ gravity is a subclass of ST gravity. The quest for the 
most general ST theory with field equations of order not higher than two has led to the 
rediscovery of Horndeski gravity \cite{Horndeski} and to its DHOST generalization which are the subject of a large literature.

Solar System  tests  constrain severely ST gravity and there 
are also stringent constraints  from Big Bang nucleosynthesis. It was proposed long ago by 
Damour and Nordvedt that ST gravity was quite different from GR early in 
the history of the universe  and that an {\em attractor-to-GR mechanism} brought it 
to GR during the radiation era, by making the scalar $\phi$ constant 
through an effective damping in its equation of motion  
\cite{Damour:1992kf,Damour:1993id}. This idea has generated a large literature with 
mixed conclusions about the presence and effectiveness of this attractor mechanism in the radiation, matter or  inflationary eras. The emerging picture is more involved than originally envisaged: the 
attractor-to-GR mechanism competes with a {\em repellor mechanism} \cite{Serna:2002fj} and it is clear that 
whether the final universe is a GR or a ST one depends heavily on the initial 
conditions, the scalar field potential $V(\phi)$ \footnote{The original 
Damour-Nordvedt scenario contained a massless scalar field with no potential.}, and its matter content. The analysis of the approach of ST 
gravity to GR is complicated by the non-linearity of the field equations.  Unfortunately,  one must make 
assumptions on the solutions themselves to draw conclusions on the dynamics of ST gravity. 
Essentially, no new result has been obtained for several years. 

We {\em revisit this subject} in the light of a new ``thermal'' view  consisting of an 
analogy between ST gravity and a dissipative fluid. The approach of ST gravity to GR is analogous  to the relaxation of a viscous fluid to its ``zero temperature'' equilibrium 
state corresponding to GR. Configurations in which the extra degree of freedom $\phi$ is 
excited correspond to positive temperature states. This new formalism introduces a 
notion of ``temperature of gravity'' and an explicit equation describing the approach to 
the GR equilibrium or its departure from it, which makes it ideally suited to address the 
issue of the attractor-to-GR mechanism  in cosmology. 
{\em This formalism, however, is not restricted to cosmology} and we first discuss it in full  
generality.

Consider first-generation ST gravity \footnote{We follow the notation 
of Ref.~\cite{Wald:1984rg}.} described by the Jordan frame action 
\be 
S_\mathrm{ST} = \int \frac{d^4 x}{16\pi} \sqrt{-g} \left[ \phi R 
-\frac{\omega}{\phi} \, \nabla^c \phi \nabla_c \phi -V \right] + 
S^\mathrm{(m)} \,, \label{action} 
\ee 
where $g$ is the determinant of the spacetime metric $g_{ab}$ with Ricci scalar $R$,  the  
``Brans-Dicke coupling'' $\omega>-3/2$ to avoid $\phi$ being  a 
phantom field, and $S^\mathrm{(m)}$ is the matter action.  The Jordan frame field equations 
are 
\begin{eqnarray}  
&& R_{ab}-\frac{R}{2} \,  g_{ab} = \frac{8\pi}{\phi} \, T_{ab}^\mathrm{(m)} + T_{ab}^{(\phi)} \,, \label{fe1}\\ 
&& \,\,\,T_{ab}^{(\phi)} := \frac{\omega}{\phi^2} \left( \nabla_a \phi \nabla_b 
\phi -\frac{1}{2} \, g_{ab} \nabla^c \phi \nabla_c \phi \right) \nonumber\\ 
&& \qquad\quad\quad +\frac{1}{\phi} \left( \nabla_a \nabla_b \phi - g_{ab} \Box\phi \right) -\frac{V}{2\phi} 
g_{ab} \, , \, \nonumber\\
&& (2\omega+3) \Box \phi = 8\pi T^\mathrm{(m)} 
+\phi \, V' -2V -\omega' \, \nabla^c\phi \nabla_c\phi 
\,,\nonumber\\
 \label{fe2} 
\end{eqnarray} 
where $R_{ab}$ is the Ricci tensor, $T_{ab}^\mathrm{(m)} $ is the matter stress-energy tensor, $T^{(m)} := g^{ab} T_{ab}^\mathrm{(m)}$, $\nabla$ is the  Levi-Civita connection, $\Box := g^{ab} \nabla_a 
\nabla_b$,  $T_{ab}^{(\phi)}$ is an effective stress-energy tensor of $\phi$, and a prime denotes differentiation   with respect to $\phi$.  

When the gradient $\nabla^c\phi$ is timelike and future-oriented,  the effective stress-energy tensor of $\phi$ describes an effective dissipative fluid with 4-velocity 
$u^a := \nabla^a \phi /\sqrt{2X}$, with $2X = -\nabla^e \phi \nabla_e\phi$. The general form of a dissipative imperfect fluid of 4-velocity $u^a$ reads 
\be
T_{ab} =\rho u_a u_b 
+P h_{ab} +\pi_{ab} + 2 \, q_{(a} u_{b)} 
\,, 
\label{eq:imperfect}
\ee
where $\rho$, $P$, $\pi_{ab}$ and $q_a$ are, respectively, an effective energy density, effective isotropic pressure, effective anisotropic stress tensor, effective heat flux density, $h_{ab} := g_{ab}+u_au_b$, and the parentheses denote the symmetrized product of $q_a$ and $u_a$. We shall denote the fluid quantities associated with $T^{(\phi)}_{ab}$ by $\rho^{(\phi)}$, $P^{(\phi)}$, $\pi_{ab}^{(\phi)}$ and $q_a^{(\phi)}$. Miraculously, the Eckart-Fourier constitutive law for dissipative fluids \cite{Eckart40} 
$ 
q_a = -{\cal K} \left( \nabla_a {\cal T} + {\cal T} \dot{u}_a \right) $ 
(where ${\cal K}$ is the thermal conductivity, ${\cal 
T}$ is the temperature, and $\dot{u}^a := u^c \nabla_c u^a$ is the fluid acceleration) 
is obeyed by this effective fluid. In fact, a  
direct computation yields $q_a^{(\phi)} =-{\cal KT} \dot{u}_a$ and  
\cite{Faraoni:2018qdr,Faraoni:2021lfc,Faraoni:2021jri,Giusti:2021sku,Gallerani:2024gdy} 
\be  
{\cal KT} = \frac{ 
\sqrt{-\nabla^c\phi \nabla_c\phi} }{ 8\pi \phi} \,. \label{KTdefinition} 
\ee 
ST gravity reduces to GR in the limit $\phi \to$~const., in which ${\cal KT} \to 0$. 
The approach/departure of ST gravity to/from GR is described by  \cite{Faraoni:2018qdr,Faraoni:2021lfc,Faraoni:2021jri,Giusti:2021sku,Gallerani:2024gdy}
\be 
\frac{d\left( {\cal KT}\right)}{d\tau} = 8\pi \left( {\cal KT}\right)^2 
-\Theta {\cal KT} +\frac{ \Box\phi}{8\pi \phi} \,,  \label{evolution_general} 
\ee 
where 
$\tau$ is the proper time along the effective fluid lines and 
$\Theta := \nabla_c u^c$ is its expansion scalar. Equation~(\ref{fe2}) yields 
\begin{widetext}
\be
\frac{d \left( {\cal KT}\right)}{d\tau} = 8\pi \left( {\cal KT}\right)^2 
-\Theta {\cal KT} + \frac{ T^\mathrm{(m)} }{\left(  2\omega + 3 \right) \phi} +\frac{1}{8\pi 
\left( 2\omega + 3 \right)} \left(  V' -\frac{2V}{\phi} -\frac{\omega'}{\phi} \, 
\nabla^c\phi \nabla_c\phi \right) \,.
\label{evolution_general2} 
\ee  
\end{widetext}
Gravity  is ``heated'' ($d\left( {\cal KT}\right)/d\tau>0$) by positive terms in 
the  right-hand side and is ``cooled'' ($d\left( 
{\cal  KT}  \right)/d\tau<0$) by negative ones. We examine the combination  ${\cal KT} 
\left( 8\pi {\cal KT}-\Theta \right)$ for electrovacuum Brans-Dicke gravity later in this Letter. The sign of  the matter contribution coincides with that of $T^\mathrm{(m)}$. 
Assuming matter to be an imperfect (i.e., with energy-momentum  tensor of the form \eqref{eq:imperfect}, see \cite{Eckart40}) or a perfect (corresponding to  
$\pi^\mathrm{(m)}_{ab} =0$, $q^\mathrm{(m)}_a=0$ and zero viscous pressure) fluid, then the trace 
is $T^\mathrm{(m)} = -\rho^\mathrm{(m)} +3 P^\mathrm{(m)}$, which is negative if $ 
P^\mathrm{(m)} < \rho^\mathrm{(m)}/3$, as guaranteed by the strong energy condition: 
then {\em matter ``cools'' gravity}. In particular, everything else equal, during the 
matter-dominated era with $T^\mathrm{(m)}=-\rho^\mathrm{(m)}$ it is 
easier for gravity to converge to GR. 

Conformal matter has $T^\mathrm{(m)}=0$ and  includes pure electromagnetic fields, 
a radiation fluid with equation of state $ P^\mathrm{(m)} =\rho^\mathrm{(m)} /3$ (e.g., in the cosmic 
radiation era) and a scalar field $\psi$ conformally coupled to the Ricci scalar $R$ 
with quartic potential $U(\psi) = \lambda \psi^4$. 
Next, it is fortunate that one can draw general conclusions on the contribution of the potential: the combination $\left( V' - 2V/\phi \right)$ gives a positive contribution, ``heating'' gravity if $ V'-2V/\phi>0$ 
or (since $ V>0 $) $  \frac{V'}{V} -\frac{2}{\phi} = \frac{d}{d\phi} 
\ln \left( \frac{V}{\phi^2} \right) >0 $.  Since the logarithm is monotonically 
increasing, this condition means that $V(\phi)$ grows faster than $\phi^2$. 
Potentials slower than $\phi^2 $ ``cool'' ST gravity, a quadratic potential does 
not affect it, and potentials faster than quadratic ``heat'' gravity, which is   particularly relevant during inflation driven by $\phi$ when matter 
is negligible.  Since $\nabla^e\phi \nabla_e \phi<0$ in  
our formalism, the last term in the right-hand side of Eq.~(\ref{evolution_general2}) 
``heats'' gravity when $\omega(\phi)$ is an increasing function and ``cools'' it when it 
is decreasing. Of course, for arbitrary functions $\omega(\phi)$  the sign of this 
contribution depends on the value of $\phi$, but the scarce literature on  ST theories 
with varying $\omega$ focuses on monotonic functions $\omega(\phi)$ and then this 
contribution has a well-defined sign. Using the definition (\ref{KTdefinition})  of ${\cal KT}$, this term is seen to be proportional to $\left( {\cal KT}\right)^2$ and Eq.~(\ref{evolution_general2}) can be rewritten as 
\be
\begin{split}
\frac{d \left( {\cal KT}\right)}{d\tau} &= 8\pi \left[ 1+  \frac{\phi}{2} \, \frac{d\ln(2\omega+3)}{d\phi} \right] \left( {\cal KT}\right)^2 -\Theta \, {\cal KT}\\ &\quad +\frac{ T^{(m)} }{\left( 2\omega + 3 \right) \phi} +\frac{V'-2V/\phi}{8\pi \left( 2\omega + 3 \right)} \,.
\label{BOH}
\end{split}
\ee
Let us examine physical situations in which Eq.~(\ref{BOH}) simplifies substantially.

\begin{figure}
    \centering
    \includegraphics[width=0.85\linewidth]{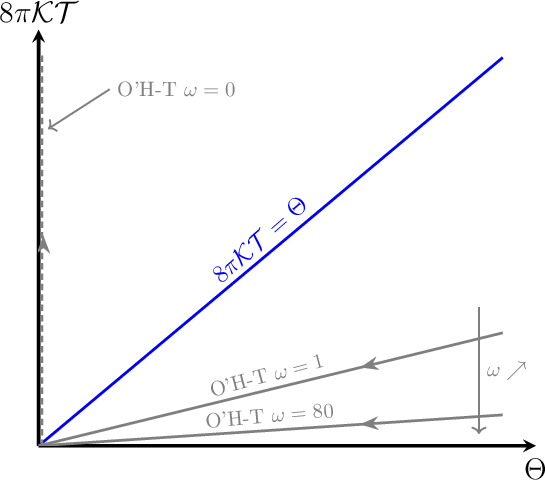}
    \caption{The critical half-line in the $\left( \Theta, {\cal KT} \right)$-plane and lines representing the cosmological O'Hanlon-Tupper solutions for two values of $\omega$. Solutions starting above this critical line deviate forever from GR, those starting below  converge to GR. The critical line cannot be crossed.}
    \label{fig:plot}
\end{figure}

\noindent {\it Electrovacuum Brans-Dicke gravity}---In Brans-Dicke gravity,  $\omega=$~const., 
$V(\phi) =0$  and we assume electrovacuum or conformal matter.
Relevant cosmological situations are the early universe in 
which $\phi$ dominates the cosmic dynamics, and the 
radiation era, but the discussion applies to general geometries and is not restricted to cosmology. Then 
$\Box \phi=0$ and Eq.~(\ref{evolution_general2}) reads
\be
\frac{d}{d\tau} \left( {\cal KT}\right) = {\cal KT} \left( 
8\pi  {\cal KT}  -\Theta  \right)  \label{simple1}
\ee
with ${\cal KT} \geq 0$. 
A partial analysis of this equation was given in~\cite{Faraoni:2021lfc, Faraoni:2021jri} but 
here we provide a much more compact and intuitive analysis noting that the $\left( \Theta, {\cal KT} \right)$-plane allows for a 
quick visualization of the dynamics of gravity (Fig.~\ref{fig:plot}), especially in regard to 
the approach of ST gravity to GR.  Although $\Theta= \nabla_c u^c = \nabla_c \left( \nabla^c\phi/ \sqrt{2X} \right)$ ultimately depends on ${\cal KT}$ through $\phi$ and its gradient, this plane is particularly convenient. It is not  a phase plane and  trajectories of the system representing the 
evolution of gravity in this plane can, in principle, intersect. The $\Theta$-axis 
corresponds to thermal equilibrium at  ${\cal KT}=0$.

By the definition~(\ref{KTdefinition}) of ${\cal KT}$, only 
the half-plane ${\cal KT} \geq 0$ is relevant. There 
are two, and only two, types of solutions with ${\cal KT}=$~constant: 
${\cal KT}=0$ and ${\cal KT}=\Theta/8\pi$. In fact, since  the 
left-hand side of 
Eq.~(\ref{simple1}) vanishes identically, it is ${\cal KT} \left( 8\pi 
{\cal KT}-\Theta \right)=0$, and they all lie on the ${\cal KT}=0$ axis 
or on the half-line  
${\cal  KT}=\Theta/8\pi$ through the origin. Since ${\cal 
KT}=$~const. for these solutions, 
also $\Theta=8\pi {\cal KT}$ is constant and this entire half-line is 
composed of points 
$ \left( \Theta , {\cal KT} \right) =  \left( \Theta_0,  8\pi \Theta_0 
\right)$ with constant $\Theta_0$. Typical solutions of this kind 
are de Sitter spaces with non-constant scalar field $\phi$, which are 
impossible in GR with a minimally coupled scalar field and are a signature 
of ST gravity. 

In the quadrant $ (\Theta >0, \, {\cal KT}>0)$ it is 
\be
\frac{d \left( {\cal KT}\right)}{d\tau} >0 \quad \Leftrightarrow \quad {\cal KT} 
>\frac{\Theta}{8\pi} \,;
\ee
if a solution of the BD field equations begins above the line ${\cal 
KT}=\Theta/8\pi$, then ${\cal KT}$ always increases (while the nearby 
solution on this line has constant ${\cal KT}$ and $\Theta$ instead) and 
${\cal KT} \to +\infty $, diverging away from GR.  If 
 a solution instead begins below this line with ${\cal KT}  < 
\Theta/8\pi$, then ${\cal 
KT}$ decreases and the solution always remains below this line, converging 
to the $ \Theta$-axis with ${\cal KT}=0$ corresponding to the GR state of 
thermal equilibrium.  

The half-line ${\cal KT}=\Theta/ 8\pi$ (referred to as the {\em critical line}) is not 
the trajectory of a solution, it  cannot be 
crossed dynamically by solutions, and it separates the $ (\Theta >0, \, {\cal KT}>0)$ 
quadrant 
into two dynamically distinct regions.  {\em A new notion of ``thermal 
stability'' of 
gravity emerges} from this picture: if the special solutions represented by 
points on the critical line are displaced slightly, 
their ${\cal KT}$ either diverges or goes to zero bringing them to GR. In 
this sense, all these special ``point-solutions'' are {\em unstable} since any 
displacement from the critical line brings them away from it (more on  
this later).

The quadrant $ (\Theta <0, \, {\cal KT}>0)$ is particularly simple (and for this reason it is not depicted in Fig.~\ref{fig:plot}): here 
$ d\left({\cal KT}\right)/d\tau = {\cal KT}\left( 8\pi {\cal 
KT}+|\Theta|\right)\geq 0$ and all solutions starting above the ${\cal 
KT}=0$ axis move upwards, diverging away from GR. This is the 
visualization of the 
sharp property of ST gravity, already enunciated in \cite{Faraoni:2021lfc, 
Faraoni:2021jri}, that the contraction 
of the 3-space seen by 
observers comoving with the effective $\phi$-fluid ``heats'' gravity. 
Therefore it is expected that, near spacetime singularities where the 
effective fluid lines converge, gravity deviates strongly from GR.  This 
is the situation, for example, of contracting universes (which ``heat'' 
away from GR) and have ${\cal KT}\to +\infty$ near Big Crunch 
singularities. 

Following Ref.~\cite{Ellis:1971pg}, we introduce  a typical length 
scale $\ell$ so that $\Theta =\frac{3}{\ell}\frac{d \ell}{d\tau} $, then $ \ell^3$ 
is the proper volume of a region of 3-space with unit  comoving 
volume and   
Eq.~(\ref{simple1}) reads 
\be
\frac{d}{d\tau}\ln \left( \ell^3 {\cal KT} \right)= 8\pi {\cal KT} >0 \,; 
\ee
$\ell^3 {\cal KT}$ cannot decrease in time, both in $\tau$-time and in 
coordinate time $t$ because their direction coincide  since 
the 4-velocity $u^a$ of the effective $\phi$-fluid is future-oriented,  $u^0 =dt/d\tau >0$. Hence, if $\ell$ decreases (i.e., 
in the $\Theta<0$ quadrant), ${\cal KT}$ increases and gravity departs 
from GR; ${\cal KT}$ can only decrease if the 3-space expands (i.e., $
\Theta>0$). 

Let us study now special solutions of ST gravity that have 
already attracted attention in the thermal view of 
scalar-tensor gravity \cite{Giardino:2023qlu}: stealth solutions and de 
Sitter spaces with non-constant scalar field.  Placing the analysis in the 
context of the $\left( \Theta, {\cal KT} \right)$-plane greatly elucidates 
their role in this formalism. 

\noindent {\it Stealth solutions}---Stealth solutions of ST gravity are vacuum solutions  
with Minkowski geometry and non-constant scalar $\phi$, for which  terms in the effective stress-energy tensor 
$T_{ab}^{(\phi)}$ conspire to cancel out, but do not vanish identically. We restrict to stealth solutions for which  $\nabla^a \phi$ is timelike and future-oriented and, in this 
section, we further require that $\Box \phi=0$ \footnote{These 
requirements exclude several legitimate  stealth solutions.}, which is satisfied if  
$\phi=\phi(t)$ and $\dot{\phi}<0$. 

In Minkowski space $\Theta = 0$ and $\Box \phi =0$ imply that
\be
\label{stealth1}
 \nabla^a\phi \nabla^b \phi \nabla_a \nabla_b \phi= 
 \dot{\phi}^2 \, \ddot{\phi} = 0 \,.
\ee
Thus, the Brans-Dicke scalar field must be 
linear \footnote{Ref.~\cite{Giardino:2023qlu} 
discusses other stealth solutions with $\phi(t) =\phi_0 t^{\beta}$ or 
$\phi(t)= \phi_0 \, \mbox{e}^{\alpha \, t}$, but these are not solutions 
of 
Brans-Dicke gravity under the assumptions made in this section.}, $ 
\phi(t)=-|\phi_0|t +\phi_1 $ 
with $\phi_{0,1}$ constants and $\phi_0<0$, $\phi_1>0$. To keep $\phi>0$, 
only the semi-infinite time interval $t< \phi_1/|\phi_0| \equiv t_*$ is 
possible. As 
$t\to t_*^{-}$, the scalar field vanishes and the effective gravitational 
coupling $G_\mathrm{eff}=1/\phi \to +\infty$. Even though the geometry 
is always exactly flat, a singularity of the gravitational coupling 
strength  
can rightly be regarded as a singularity of gravity. 

In the $\left( \Theta, {\cal KT} \right) $-plane, stealth solutions live on the vertical ${\cal K} {\cal T}$-axis and can only 
move upward in time because $d\left( {\cal KT} \right)/d\tau=8\pi \left( 
{\cal KT} \right)^2  >0$. This equation has the pole-like solution 
\be
{\cal KT} = \frac{1}{ 8\pi \left| \tau_0-\tau\right| } \label{stealth2}
\ee
exploding in a  finite time $\tau_0$ 
(only the increasing branch has physical meaning). All these stealth solutions run infinitely 
far away from GR in a finite time.

\noindent {\it de Sitter solutions with non-static scalar field}---A signature of ST gravity is the existence of vacuum de Sitter 
solutions with non-constant scalar field. By contrast, in GR sourced by a 
minimally coupled scalar field, de Sitter spaces 
necessarily have constant scalar field. 

In the $\left( \Theta, {\cal KT} \right)$-plane, these de Sitter solutions correspond to points on the critical line 
with $\Theta=3H=$~const., where $H\equiv \dot{a}/a$ is the Hubble function 
of the FLRW line element
\be
ds^2 =-dt^2 +a^2(t) \left( dx^2+dy^2+dz^2 \right) \,.\label{FLRW}
\ee
If $H=0$ the de Sitter geometry degenerates into the Minkowski one and we 
recover the $\Theta=0$ stealth solution~\eqref{stealth1}, 
\eqref{stealth2}. The equation $\Box\phi=-\left( \ddot{\phi}+3H \dot{\phi} \right) =0 $  admits the first integral $\dot{\phi}= C/a^3$, where $C$ is a negative constant (for future-orientation), therefore $ \phi=(|C|a_{0}/\Theta)\,{\rm e}^{-\Theta\,t}$ and ${\cal KT}= |\dot{\phi}|/ (8\pi \phi)=\Theta/8\pi$, so these de Sitter spaces are points lying on the critical line of the $\left( \Theta, {\cal KT} \right)$-plane (most of the de Sitter solutions with non-static $\phi$  in~\cite{Giardino:2023qlu} have $\Box \phi\neq 0$).
%
\begin{comment}
{\red the only ones with 
$\Box\phi=0$ have 
\be
H=H_0=\mbox{const.}, \quad\quad \phi(t)= \phi_0 \, \mbox{e}^{\alpha t} \,, 
\quad \alpha<0\,,
\ee
with $\phi_0 $ a positive constant. The condition $\Box\phi=-\left( 
\ddot{\phi}+3H_0 \dot{\phi} \right)=0$ requires $\alpha=-3H_0<0$. Then \cite{Giardino:2023qlu},
\be
{\cal KT}= \frac{ |\alpha|}{8\pi} = \frac{\Theta_0}{8\pi} 
\ee
and these solutions are points lying exactly on the critical line.  They 
are found to be unstable with 
respect to perturbations in \cite{Giardino:2023qlu}, although the 
stability criterion is different from our considerations in the present 
paper.}
\end{comment}
No other de Sitter solutions with non-static $\phi$ are found under the 
restrictions discussed so far.

\vfill\null

\noindent {\em O'Hanlon and Tupper (O'H-T) solution.}---As an explicit example, let us consider the O'H-T solution of ST cosmology \cite{OHanlon:1972ysn}, corresponding to the theory with $V=0$, $T_{ab}^{(m)}=0$, $\omega>-3/2$ and $\omega\neq-4/3$. The solution is represented by power laws for the scale factor and scalar field, $a\sim t^{q_{\pm}}$ and $\phi\sim t^{s_{\pm}}$, with
${q_{\pm}}=\omega/[3(\omega+1)\mp\sqrt{3(2\omega+3)}]$ and ${s_{\pm}}=1-3q_{\pm}$. Requiring $\phi>0$ and $\nabla^a \phi$ future-oriented let us select the $(+)$ solution, for which we have \cite{Giardino:2022sdv} $8\pi{\cal KT}= |s_{+}|/t$. Furthermore, since $\Theta = 3H = 3q_{+}/t$, the O'H-T solution corresponds to the straight line
$$
8\pi\,{\cal KT}=\frac{|s_{+}|}{3q_{+}}\,\Theta = \left|\frac{1}{3q_{+}}-1\right|\,\Theta \, ,
$$
which implies that $8\pi\,{\cal KT}<\Theta$ for all $\omega>-3/2$, $\omega\neq-4/3 \, , \, 0$. Hence, based on our argument, gravity in the O'H-T solution, with $\omega>-3/2$, $\omega\neq-4/3 \, , \, 0$, must relax to GR, which is exactly what happens at late times. Furthermore, the case $\omega=0$ corresponds to $a\sim 1$ and $\phi \sim -t$ with $t<0$. Then $\Theta = 0$ and $8\pi\,{\cal KT} = 1/|t|$ is above the critical line. Hence, as $t$ increases, gravity diverges away from GR. These scenarios are represented in Fig. \ref{fig:plot},  where the arrows denote the behavior of ${\cal KT}$ as $t$ increases.

\noindent {\it Attractor-to-GR mechanism}---With the assumptions made in this section, we are able to assess the attractor-to-GR mechanism found by Damour and Nordvedt \cite{Damour:1992kf,Damour:1993id} and discussed by many authors, including  i)~an early epoch in which a free Brans-Dicke field $\phi$ dominates the cosmic dynamics and matter is irrelevant; ii)~the radiation epoch in which $T^\mathrm{(m)}=0$ and $\phi$ is massless (which is the original situation of \cite{Damour:1992kf,Damour:1993id}).

In both cases we assume a spatially flat  FLRW universe with line 
element~(\ref{FLRW}) \footnote{It is known that spatial curvature affects the 
attractor-to-GR mechanism \cite{Santiago:1998ae}. Spatially curved universes will be 
discussed elsewhere.} and Brans-Dicke field $ \phi(t)$. 
In both situations $ \ddot{\phi}+3H\dot{\phi} =0$ and the comoving time $t$ coincides with the proper time 
$\tau$ of the effective $\phi$-fluid. 

In the light of the previous discussion, the attractor-to-GR mechanism 
operates if and only if ${\cal 
KT}<\Theta/(8\pi)$ initially. In a FLRW universe 
$\Theta=3H=3\dot{\ell}/\ell$ 
and $\ell=a$ for a unit comoving volume. Using 
the definition~(\ref{KTdefinition}) of ${\cal KT}$, this criterion reads
\be
\sqrt{-\nabla^e\phi \nabla_e\phi} < 3H\phi \,,
\ee
which is never satisfied in a  contracting universe with $H<0$.  In an expanding universe, this criterion becomes
$   |\dot{\phi}| /\phi <  3H $, 
which reads as: there is an attractor-to-GR mechanism if and only if the 
scale of variation of the scalar field $\tau_{\phi} \equiv 
\phi/|\dot{\phi}|$ 
(equivalently,  the scale of variation $\tau_G=\tau_{\phi} := 
G_\mathrm{eff}/ |\dot{G}_\mathrm{eff}|$) satisfies
\be
\tau_{\phi}> 	\frac{\tau_H}{3} := \frac{1}{3H} \,.\label{rates}
\ee 
The physical interpretation is that if $\phi $ varies too fast (${\cal 
KT}$ above the critical line), GR is never approached while, if it 
varies sufficiently slowly, GR is always approached. The 
effective strength of gravity is precisely the extra 
degree of freedom added to the two massless spin-2 modes of GR. When the  
variation of $\phi$ is ``fast'', the scalar mode dominates over the two GR modes 
while the opposite is true when $\phi$ and $G_\mathrm{eff}$ vary slowly 
and the GR modes dominate, in a way made precise by Eq.~(\ref{rates}) or 
by the thermal criterion ${\cal KT} < \Theta/8\pi$. Keep in mind that 
the 
scenario applies to the cosmic radiation era even though the scalar field 
is not dominant.  Thus, the thermal view of scalar-tensor gravity provides insight on the 
attractor-to-GR problem in the restricted context of Brans-Dicke gravity 
with $\Box\phi=0$. The results can be extended  immediately to the case  $V(\phi)= m^2 \phi^2/2$, the potential that         does not contribute to $\Box \phi$ (it does, however, contribute to 
sourcing $g_{ab}$, therefore the simple stealth and de Sitter solutions  already discussed are no longer valid). 

\noindent {\it Conclusions}---For an expanding 3-space (i.e., $\Theta>0$) and $\Box\phi=0$, Brans-Dicke gravity ``cools'' 
and converges to GR if 
it starts sufficiently close to it (i.e., below the critical line), while it ``heats up'' and runs away 
from GR if it is sufficiently ``hot'' (i.e., above the critical line, therefore sufficiently far away from GR) initially. These conclusions hold, in particular, in FLRW cosmology and explain the attractor-to-GR mechanism and the repellor mechanism discovered in previous literature, without assuming anything on the solutions themselves, apart from the initial position in the $\left( \Theta, {\cal KT}\right)$-plane. A clear physical interpretation of why the scalar degree of freedom dominates over the two tensor ones emerges naturally from this description.

Contrary to  Refs.~\cite{Damour:1992kf,Damour:1993id} and to most authors, who work in the Einstein conformal frame, we have performed our analysis entirely in the Jordan frame. However, the thermal analogy of scalar-tensor gravity has an equivalent description in the Einstein frame, in which one trades temperature (which is always zero) with chemical potential $\mu$ and an evolution equation for $\mu$ analogous to~(\ref{evolution_general})  is present \cite{Faraoni:2022gry}.

Here we have covered the original scenario of \cite{Damour:1992kf,Damour:1993id}. More involved situations in which $\Box\phi \neq 0$ and the extra terms in the right-hand side of Eq.~(\ref{BOH}) play a role will be discussed in future work, in the context of scenarios  proposed in the literature and involving specific potentials, matter fluids and $\phi$-dependent Brans-Dicke couplings. However, we can anticipate two results that highlight the generality of our approach compared to the current literature. Within the assumptions of the thermodynamics of scalar-tensor gravity, if $V>0$, $V$ grows faster than $\phi^2$, $2\omega+3> 0$, $\omega'\geq 0$, and assuming a matter perfect fluid with $w=p/\rho > 1/3$, then the last two terms in Eq.~\eqref{evolution_general2} are positive and
$$\frac{d \left( {\cal KT}\right)}{d\tau} \geq 8\pi \left( {\cal KT}\right)^2 
-\Theta {\cal KT} \, ,$$
hence if ${\cal KT}>\Theta/8\pi$ the theory will diverge away from GR also for more general models including matter and non-trivial potentials. Viceversa, for a perfect fluid $w < 1/3$, $2\omega+3> 0$, $\omega'< 0$, and $V$ growing slower than $\phi^2$, then
$$\frac{d \left( {\cal KT}\right)}{d\tau} < 8\pi \left( {\cal KT}\right)^2 
-\Theta {\cal KT} \, ,$$
which guarantees the existence of an attractor to GR if ${\cal KT}<\Theta/8\pi$. In other words, the first scenario pushes the critical curve separating the two regions in the $(\Theta,{\cal K}{\cal T})$-plane down toward the $\Theta$-axis, whereas the second scenario pushes it up above the critical line ${\cal K} {\cal T} = \Theta/8\pi$. More involved scenarios, to further extend the generality of this analysis, will be discussed elsewhere.

\bigskip

V.F. is supported by the Natural Sciences \& Engineering Research Council 
of Canada (Grant no. 2023-03234). A.G.~is supported in part by the Science and Technology Facilities Council (grants n.~ST/T006048/1 and ST/Y004418/1). The work of A.G. has been carried out in the framework of activities of the National Group of Mathematical Physics (GNFM, INdAM).   

\bibliography{PlaneBib}

\end{document}